\documentclass[preprint,pra,onecolumn]{revtex4}
\usepackage{amsfonts}
\usepackage{amsmath}
\usepackage{amssymb}

\usepackage{fixme}
\usepackage{graphicx}
\usepackage{dcolumn}
\usepackage{bm}
\usepackage{SIunitx}
\usepackage{color}

\begin{document}
\title{An efficient implementation of the decoy-state measurement-device-independent quantum key distribution with heralded single-photon sources}
\author{Qin Wang$^{1}$}
\email{qinw@njupt.edu.cn}
\author{Xiang-Bin Wang$^{2,3}$}
\email{xbwang@mail.tsinghua.edu.cn}
\affiliation{$^{1}$Institute of Signal Processing and Transmission, Nanjing University of Posts and Telecommunications, Nanjing 210003, China}
\affiliation{$^{2}$Department of Physics and State Key Laboratory of Low Dimensional Quantum Physics, Tsinghua University, Beijing 100084, China}
\affiliation{$^{3}$Jinan Institute of Quantum Technology, Shandong Academy of Information Technology, Jinan, China}

\begin{abstract}
 We study the decoy-state measurement-device-independent quantum key distribution using heralded single-photon sources. This has the advantage that the observed error rate in $X$ basis is in higher order and not so large. We calculate the key rate and transmission distance for two cases: one using only triggered events, and the other using both triggered and non-triggered events. We compare the key rates of various protocols and find that our new scheme using triggered and non-triggered events can give higher key rate and longer secure distance. Moreover, we also show the different behavior of our scheme when using different heralded single-photon sources, i.e., in poisson or thermal distribution. We demonstrate that the former can generate a relatively higher secure key rate than the latter, and can thus work more efficiently in practical quantum key distributions.

PACS number(s): 03.67.Dd, 03.67.Hk,42.65.Lm

\end{abstract}

\maketitle

\section{Introduction}

As is well known that the quantum key distribution (QKD) is standing out compared with conventional cryptography due to its unconditional security based on the law of physics. It allows two legitimate users, say Alice and Bob, to share secret keys even under the present of a malicious eavesdropper, Eve. But its security proofs often contain certain assumptions either on the sources or on the detection systems, and usually practical setups have imperfections. Therefore, the "in-principle" unconditional security can actually conflict with realistic implementations, and which might be exploited by Eve to hack the system \cite{PNS1,PNS2,Fung,Lyde}.

In order to achieve the final goal of unconditional security in practice, different approaches have been proposed, such as the decoy-state method \cite{gott1,gott2,hwan,wang1,lo1}, the device-independent quantum key distribution (DI-QKD) \cite{Maye,Gisi} and recently the measurement-device-independent quantum key distribution (MDI-QKD) \cite{lo2,Brau}. Among them, the decoy-state MDI-QKD seems to be a promising candidate considering its relatively lower technical demanding.

The decoy-state MDI-QKD was studied extensively with infinite different intensities\cite{lo2} and a few intensities\cite{wang2}. However, the efficient decoy-state MDI-QKD with heralded source is not shown. We know that weak coherent states (WCSs) at least have two drawbacks: one is the large vacuum component, and the other is the significant multi-photon probabilities. The former leads to a rather limited transmission distance, since the dark count contributes lots of bit-flip errors for long distance. The latter one results in a quite low key generation rate. In the existing MDI-QKD \cite{lo2,Brau} setup, all detections are done in Z basis. There are events of two incident photons presenting on the same side of the beam-splitter and no incident photon on another side. Such a case can cause a quite high observed error rate in X basis. Though in principle one can deduce the phase-flip error rate by comparison of the observed error rate in X basis for different groups of pulses as shown in \cite{wang2}, the high error rate in X basis can still decrease the key rate drastically in real implementations when we take statistical fluctuations into account. Fortunately, besides the WCSs, there is another practically easy implementable source, the heralded single-photon source (HSPS). The source can eliminate those drawbacks, and give much better performance than WCSs in the QKD \cite{qin1,qin2}, since the dark count can be eliminated to a negligible level for a triggered source. Moreover, the cause of a high error rate in X basis does not exist for a HSPS due to a high order small probability for events of two photons present on the same side of the beam-splitter.

We also note that it is impossible to use infinite number of decoy states in a realistic MDI-QKD, therefore, people often use one or two decoy states to estimate the behavior of the vacuum, the single-photon and the multi-photon states \cite{wang1,qin1}.

Here in this work, we study MDI-QKD with heralded single-photon sources. We use both the triggered and non-triggered events of HSPSs to precisely estimate the lower bound of the two-single-photon contribution ($Y_{11}$) and the upper bound of the quantum bit-error rate (QBER) of two-single-photon pulses ($e_{11}$). As a result, we get an much longer transmission distance and a much higher key generation rate compared with existing decoy-state MDI-QKD methods \cite{wang2}, and come close to the result of infinite different intensities. After presenting the schematic set-up of the method, we shall  present formulas such as $Y_{11}$ and $e_{11}$ for calculating the key rate in Sec. II. In Sec. III, we proceed numerical simulations with practical parameters and compare with existing schemes. Finally, we give conclusions in Sec IV.

\begin{figure}[ptb]
\begin{center}
\includegraphics[scale=0.6]{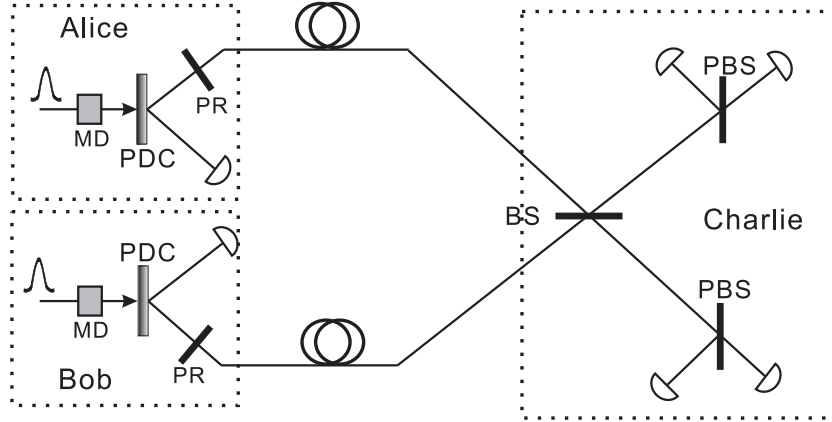}
\end{center}
\caption{(Color online)  (a). A schematic setup of the method. Alice and Bob randomly prepare HSPSs from PDC processes in a BB84 polarization state with a polarization rotator (PR). Decoy states are generated by changing the power of each pump laser with a modulator (MD). Signal pulses from Alice and Bob interfere at a 50/50 beam-splitter (BS) and then each enter a polarizing beam splitter (PBS) projecting the input photons into either horizontal (H) or vertical (V) polarization states. Four single-photon detectors are employed at the third party, Charlie's side to detect the results. Moreover, both the triggered and non-triggered events at Alice and Bob's side are sent to Charlie, and corresponding counting rates are recorded individually.}
\label{Fig1}

\end{figure}
\section{Improved method of decoy-state MDI-QKD with heralded source}
\subsection{The  method and formulas}
We know that the state of a two-mode field from the parametric-down conversion (PDC) source is \cite{yurk,lutk1}:
\begin{align*}
\left \vert \Psi \right \rangle _{TS}  &  =\sum\limits_{n = 0}^\infty  {\sqrt {P_n } } \left| n \right\rangle _T \left| n \right\rangle _S
\\
P_{n}(x)  &  = \frac{x^{n}}{\left(  1+x\right)  ^{n+1}},     (\Delta t_c  \gg \Delta t)
\end{align*}
or \begin{align*} P_{n}(x)  & = e^{ - x} \frac{{x^n }}{{n!}},     (\Delta t_c  \ll \Delta t) \end{align*}
where $\left \vert n\right \rangle$ represents an $n$-photon state, and $x$ is the intensity (average photon number) of one mode. Mode T (trigger) is detected by Alice or Bob, and mode S (signal) is sent out to the untrusted third party (UTP). $\Delta t_c$ is the coherence time of the emission, and $\Delta t$ is the duration of the pump pulse. As demonstrated in \cite{mori,ribo}, we can either get a thermal distribution or a poisson distribution by adjusting the experimental conditions, e.g. changing the duration of the pump pulses. Below, we will at first use HSPSs with poisson distributions as an example to describe our new MDI-QKD scheme, and then compare it with the case of with thermal distributions.

We denote $q_n^\upsilon$ as the probability of triggering at Alice or Bob's detector when $n$-photon state is emitted,
\begin{align*}
q_0^v = d_\upsilon
\end{align*}
and
\begin{align*}
q_n^\upsilon = 1 - (1-d_v)(1 - \eta _\upsilon )^n,
\end{align*}
for $i \geqslant 1$. Here $\upsilon$ can be A (Alice) or B (Bob), $\eta _\upsilon$ and $d _\upsilon$ are the detection efficiency and the dark count rate at Alice (Bob)'s side, respectively. For simplicity, we may omit the superscript or subscript $v$  latter if there is no confusion. Then the corresponding non-triggering probability is $(1 - q_n)$.

We request Alice (or Bob) to randomly change the intensity of her (or his) pump light among three values, so that the intensity of one mode is randomly changed among $0$, $\mu_A$ (or $\mu_B$), and $\mu_A^{\prime}$ (or $\mu_B'$) (and $\mu_A<\mu_A^{\prime},\; \mu_B<\mu_B^{\prime}$).  We define the subclass of source pulses that Alice uses intensity $x$, Bob uses intensity $y$ as {\em source} $\{x,y\}$,  $x\in \{0,\mu_A,\mu_A'\}$ and $y\in \{0,\mu_B,\mu_B'\}$. After triggering detection, there are 4 classes of states at each side from the two-mode fields of two different intensities, as there are triggered and non-triggered states from each intensity. In principle, here we have many choices in implementing the decoy-state MDI-QKD. For example, using all events in both intensities; using only triggered events of them; or using triggered events in one intensity and non-triggered events in another. Here we shall study the following two cases: 1) using only triggered events in both intensities; 2) using non-triggered events from the stronger field and triggered events from the weaker field for the estimation of $Y_{11}$, and then using the triggered events from the stronger pulses for the final key distillation. We declare that: Firstly, both cases can lead to a longer transmission distance than that of using WCSs; Secondly, both the key rate and the secure transmission distance in the second case are better than that in the first case.

As shown in \cite{lo2}, we use the rectilinear basis ($Z$) as the key generation basis, and the diagonal basis ($X$) for error testing only. We denote $Y_{mn}^{W,t}$, $S_{mn}^{W ,t}$, and $e_{mn}^{W ,t}$ to be the yield, the gain and the QBRE of the triggered signals respectively, where $n$, $m$ represent the number of photons sent by Alice and Bob, and $W$ represent the $Z$ or $X$ basis.   Similarly, we also define $Y_{mn}^{W ,nt}$, $S_{mn}^{W,nt}$, and $e_{mn}^{W,nt}$ as corresponding values for the non-triggered events.
Note that the gain $S_{x,y}^{W, t}$ is defined as $n_{x,y}^{W,t} /N_{x,y}^W$, if $n_{x,y}^{W,t}$ and $N_{x,y}^W$ are the number of {\em detected} events after triggering at both side and the number of total events (no matter triggered or not) among the subclass of source pulses that Alice uses intensity $x$, Bob uses intensity $y$ and both of them are prepared in basis $W$. Similar definition is also used for $S_{x,y}^{W ,nt}$, the gain of non-triggered sources in basis $W$. All gains can be directly experimentally observed, and thus are regarded as {\em known} values. The yield $\{Y_{mn}^{W, t}\}$ is defined as the the rate of producing a successful event for two-pulse state $|m\rangle\otimes|n\rangle$ prepared in $W$ basis after triggering. Similar definition is also used for non-triggered pulses. Asymptotically, we have $Y_{mn}^{W,t}=Y_{mn}^{W,nt}$. Therefore we shall only use $Y_{mn}^{W}$ for both of them. Note the the yield of $Y_{mn}^{W}$ is not directly observed in the experiment and our first major task is to deduce the lower bound of $Y_{11}^W$ based on the known values, $\{S_{xy}^{W,t}\}, \{S_{xy}^{W,nt}\}$. Here we assume to implement the decoy-state method in different bases {\em separately}, therefore we shall omit the superscript $W$ here after provided that this does not make any confusion.

The un-normalized density matrix for a triggered event from intensity two-mode field of intensity $r$ is
\begin{equation}\label{xt}
\rho_{r}^t = \left({\sum\limits_0^\infty  {q_nP_n(r) \left| n \right\rangle \left\langle n \right|}} \right).
\end{equation}
Also, we have the following density matrix for a non-triggered event at Alice's side
\begin{equation}\label{xnt}
\rho_{r}^{nt} = \left({\sum\limits_0^\infty  {(1-q_n)P_n(r) \left| n \right\rangle \left\langle n \right|}} \right).
\end{equation}
Using conclusions in Ref. \cite{wang2}, we can obtain the yield of single-photon pairs once we know the source state.
For triggered events, we have
\begin{equation}\label{t1}
\begin{gathered}
  S_{x,y}^{t} = \tilde S_{00}^{t}  + \eta _A \eta _B xe^{ - x} ye^{ - y} Y_{11} + \eta _A xe^{ - x} \sum\limits_{n = 2}^\infty {[1 - (1 - \eta _B )} ^n ]e^{ - y} \frac{{y^n }}
{{n!}}Y_{1n} + \eta _B ye^{ - y} \sum\limits_{m = 2}^\infty  {[1 - (1 - \eta _A )} ^m ]  \\
e^{ - x} \frac{{x^m }}{{m!}}Y_{m1}
   + \sum\limits_{m = 2,n = 2}^\infty  {e^{ - x} \frac{{x^m }}
{{m!}}} e^{-y} \frac{{y^n }}
{{n!}}[1 - (1 - \eta _A )^m ][1 - (1 - \eta _B )^n ]Y_{mn}.  \\
\end{gathered}
\end{equation}
Here $\tilde S_{00}^{t} = \mathcal{L}_A+\mathcal{L}_B-\mathcal{L}_0$, and $\mathcal{L}_A = d_B e^{-y}\sum\limits_{m = 0}^\infty{[1-(1-d_A)(1-\eta _A)}^m]e^{-x}\frac{{x^m}}{{m!}}Y_{m0}$, $\mathcal{L}_B = d_A e^{-x}\sum\limits_{n = 0}^\infty{[1-(1-d_B)(1-\eta _B)}^n]e^{-y}\frac{{y^n}}{{n!}}Y_{0n}$, $\mathcal{L}_0 = d_A d_B e^{-x}e^{-y}Y_{00}$. According to the definition of the gains above, one easily finds that fact: ${\mathcal L}_A=S_{x0}^{t}, \mathcal{L}_B=S_{0y}^{t}, \mathcal{L}_0=S_{00}^{t}$. All these gains are known values. Therefore, $\tilde S_{00}^{t}=S_{x0}^{t}+ S_{0y}^{t}-S_{00}^{t}$ is also a {\em known} value. Similarly, we also have the following equation for the non-triggered events:
\begin{equation}\label{n1}
\begin{gathered}
  S_{x,y}^{nt} = \tilde S_{00}^{nt} + (1 - \eta _A )(1 - \eta _B )xe^{ - x} ye^{ - y} Y_{11} + (1 - \eta _A )xe^{ - x} \sum\limits_{n = 2}^\infty  {(1 - \eta _B )} ^n e^{ - y} \frac{{y^n }}
{{n!}}Y_{1n} + (1 - \eta _B )ye^{ - y}   \\
\sum\limits_{m = 2}^\infty {(1 - \eta _A )} ^m e^{ - x} \frac{{x^m }}
{{m!}}Y_{m1} + \sum\limits_{m = 2,n = 2}^\infty {e^{ - x} \frac{{x^m }}
{{m!}}} e^{ - y} \frac{{y^n }}
{{n!}}(1 - \eta _A )^m (1 - \eta _B )^n Y_{mn}.  \\
\end{gathered}
\end{equation}
where $\tilde S_{00}^{nt} = S_{x0}^{nt} + S_{0y}^{nt}  - S_{00}^{nt}$. And also $S_{x0}^{nt} = (1-d_B)e^{ - y}\sum\limits_{m = 0}^\infty {[(1 - d_A)(1 - \eta _A)}^m]e^{-x}\frac{{x^m}}{{m!}}Y_{m0}$, $S_{0y}^{nt} =(1-d_A) e^{-x}\sum\limits_{n = 0}^\infty {[(1 - d_B)(1 - \eta _B)}^n]e^{-y}\frac{{y^n}}{{n!}}Y_{0n}$, $S_{00}^{nt} = (1 - d_A )(1 - d_B )e^{-x}e^{-y}Y_{00}$. And they are regarded as known values. Now let's use $S_{\mu,\mu }^{t}$ and $S_{\mu ',\mu '}^{nt}$ to estimate a tight bound of $Y_{11}$.
Denoting $k = \frac{{(1 - \eta _A)(1 - \eta _B)^2 }}{{\eta _A [1 - (1 - \eta _B )^2 ]}}(\frac{{\mu '}}{\mu })^3 e^{2\mu - 2\mu '}$, and combining Eq. (\ref{n1}) and (\ref{t1}), we obtain
\begin{equation}\label{cond}
Y_{11}
=\frac{k (S_{\mu,\mu}^{t} - \tilde S_{00}^{t} ) - (S_{\mu ',\mu '}^{nt} - \tilde S_{00}^{nt} ) + \mathcal{K}}{[k \eta _A \eta _B \mu ^2 e^{ - 2\mu } - (1 - \eta _A )(1 - \eta _B )\mu '^2 e^{ - 2\mu '}]}
\end{equation}
and
\begin{equation}
\begin{gathered}
  \mathcal{K}=\sum\limits_{n = 2}^\infty { \{(1 - \eta _A )\mu 'e^{ - 2\mu '} (1 - \eta _B )^n \frac{{\mu '^n }}{{n!}} - k \eta _A \mu e^{ - 2\mu } [1 - (1 - \eta _B )^n] \frac{{\mu ^n }}{{n!}} \}} Y_{1n}   +   \\
  \sum\limits_{m = 2}^\infty  { \{ (1 - \eta _B )\mu 'e^{ - 2\mu '} (1 - \eta _A )^m \frac{{\mu '^m }}{{m!}} - k \eta _B \mu e^{ - 2\mu } [1 - (1 - \eta _A )^m ] \frac{{\mu ^m }}{{m!}} \} } Y_{m1}   +   \\
  \sum\limits_{m = 2,n = 2}^\infty { \{ (1 - \eta _A )^m (1 - \eta _B )^n e^{ - 2\mu '} \frac{{\mu '^m }}{{m!}} \frac{{\mu '^n }}{{n!}} - k[1 - (1 - \eta _A )^m ][1 - (1 - \eta _B )^n] e^{ - 2\mu } \frac{{\mu ^m }}{{m!}}\frac{{\mu ^n }} {{n!}} \} } Y_{mn}.    \\
\end{gathered}
\end{equation}
To lower bound $Y_{11}$ here, we can choose to set the following simultaneous conditions:
\begin{equation}\label{cond3}
[k \eta _A \eta _B \mu ^2 e^{ - 2\mu } - (1 - \eta _A )(1 - \eta _B )\mu '^2 e^{ - 2\mu '}]\le 0;\; \mathcal{K} \le 0
\end{equation}
When both conditions above are met, we have the following inequality for the lower band of $Y_{11}$:
\begin{equation}\label{y11}
Y_{11} \geqslant {Y_{11} }^L  \equiv \frac{k (S_{\mu,\mu}^{t} - \tilde S_{00}^{t} ) - (S_{\mu ',\mu '}^{nt} - \tilde S_{00}^{nt} ) }{[k \eta _A \eta _B \mu ^2 e^{ - 2\mu } - (1 - \eta _A )(1 - \eta _B )\mu '^2 e^{ - 2\mu '}]}.
\end{equation}\label{cond4}
(since the value of $\mu$ and $\mu '$ can be chosen separately, the above conditions can be easily satisfied in practice,)
In particular, in the symmetric case that $\eta_A=\eta_B=\eta$, the conditions on Eq. (\ref{cond3}) reduce to
\begin{equation}\label{sc}
\mu \ge (1-\eta)\mu'.
\end{equation}
For simplicity, we shall use such a condition for all calculations.
Actually, directly applying Eq. (16) and Eq. (2) of Ref. \cite{wang2} together with Eq. (\ref{xt},\ref{xnt}) here can also lead to Eq. (\ref{y11},\ref{sc}).
Then the gain of the two-single-photon pulses for the triggered and non-triggered signals ($\mu '$) are:
\begin{equation}
S_{11}^{t}  = \eta ^2 \mu '^2 e^{ - 2\mu '} Y_{11},
\end{equation}
\begin{equation}
S_{11}^{nt}  = (1 - \eta )^2 \mu '^2 e^{ - 2\mu '} Y_{11}.
\end{equation}

As mentioned above, we use two bases in this protocol, i.e., the Z basis and the X basis. We use the former to generate real keys, and the latter only for error test. After error test, we get the bit-flip error rates for the triggered and non-triggered pulses as $E_{\mu ,\mu }^{t}$
and $E_{\mu ',\mu '}^{nt}$. In order to calculate the final key rate, we also need to know the phase-flip error rate of two-single-photon pulses in the Z basis, i.e. $e_{11}^{ph,t}$ (or $e_{11}^{ph,nt}$) which is equal to the bit-flip rate in the X basis, $e_{11}^{X,t}$ (or $e_{11}^{X,nt}$), whose values can be represented as:
\begin{equation}
e_{11}^{X,t}  \leqslant \frac{{E_{\mu ,\mu }^{X,t} S_{\mu ,\mu }^{X,t}  - E_{\mu ,0}^{X,t} S_{\mu ,0}^{X,t}  - E_{0,\mu}^{X,t} S_{0,\mu}^{X,t}+ E_{0,0}^{X,t} S_{0,0}^{X,t} }}
{{S_{11}^{\omega ,t} }} \equiv e_a^X,
\end{equation}
or \begin{equation}
e_{11}^{X,nt}  \leqslant \frac{{E_{\mu ',\mu '}^{X,nt} S_{\mu ',\mu '}^{X,nt}  - E_{\mu ',0}^{X,nt} S_{\mu ',0}^{X,nt} - E_{0,\mu'}^{X,nt} S_{0,\mu '}^{X,nt}  + E_{0,0}^{X,nt} S_{0,0}^{X,nt} }}
{{S_{11}^{\omega ,nt} }} \equiv e_b^X.
\end{equation}
Combing the two bounds, we have \cite{note}:
\begin{equation}
e_{11}^X  \leqslant e_{11}^{X,U}  = \min \{ e_a^X ,e_b^X \}.
\end{equation}

Now we can calculate the final key generation rate for the triggered signal pulses ($\mu '$) as:
\begin{equation}\label{kr}
R^t  \geqslant \{ q^2 P_1^2(\mu') Y_{11}^{Z,t} [1 - H_2 (e_{11}^X )] - S_{\mu ',\mu '}^{Z,t} f(E_{\mu ',\mu '}^{Z,t} )H_2 (E_{\mu ',\mu '}^{Z,t} )\},
\end{equation}
where  $f(E_{\mu^{\prime}})$ is a factor for the cost of error correction given existing error correction systems in practice, and we assume $f=1.16$ here \cite{lo2}. $H_{2}\left(  x\right)$ is the binary Shannon information function, given by
\[
H_{2}\left(  x\right)  =-x\log_{2}(x)-(1-x)\log_{2}(1-x).
\]
We have not considered the effects of bases mismatch in BB84 protocol. Actually, one can choose basis in a biased way \cite{aya} and the effect can vanish asymptotically.
In fact, the non-triggered events and the triggered events from weaker fields can also be used to distill secret keys as shown in \cite{adac}. However, for simplicity, in the following simulations we consider only the triggered components from the stronger field.

\section{Numerical simulation}
With formulas above, we can now numerically calculate the key rate and compare the secret key generation rate of our new MDI-QKD scheme with existing methods \cite{lo2,wang2}. Moreover, we will show the different results of our proposed scheme using different HSPSs, i.e., in poisson or thermal distributions. Below for simplicity, we assume that the UTP locates in the middle of Alice and Bob, and the UTP's detectors are identical, i.e., they have the same dark count rate and detection efficiency, and their detection efficiency does not depend on the incoming signals.

We shall estimate what values would probably be observed for the gains and error rates in the normal cases by the linear model \cite{lo2,qinwang2} where state $|n\rangle\langle n|$ from Alice changes to
\begin{equation}
\sum_{k=0}^n C_n^k  \eta^k(1-\eta)^{n-k} |k\rangle\langle k|,
\end{equation}
where $\eta$  is the transmittance from Alice to the UTP. Using this model, we can set values (probably would-be observed values in experiments) for $S_{xy}^{t}$, $S_{xy}^{nt}$, $E_{xy}^{t}$ and $E_{xy}^{nt}$ according to transmission distance. After setting these values, we can find the distance dependent key rate by Eq. (\ref{kr}). For this purpose, we have developed a general model to simulate the probably observed gains and error rates and hence the final key rate under linearly loss channel, given whatever source state\cite{qinwang2}.

For a fair comparison, we use the same parameters as in \cite{lo2,ursi}, see Table I, except that Alice (Bob) uses an extra detector for heralding signals with a detection efficiency of $\eta_{A}$ ($\eta_{B}$) and dark count rate of ${d_A}$ (${d_B}$).

In practical implementations, people often use a non-degenerate PDC process and obtain a visible and a telecommunication wavelength in mode T and S, respectively. To simplify the simulations, we assume both Alice and Bob have the same silicon avalanche photodiodes. We do calculation for the conditions of detection $\eta_{A}=\eta_{B}=0.75$ (or $0.9$), and $d_{A}=d_{B}=10^{-6}$. At each distance, in order to maximize the key generation rate, we set $\mu=(1-\eta)\mu'$ and use the optimal $\mu'$ for the case of using both triggered and non-triggered events, for other cases we set $\mu=0.1$ and use the optimal value of ${\mu '}$. Our simulation results are shown in Figs. 2 - 4.

\begin{table}[ptb]
\caption{Parameters used in numerical simulations: $\alpha$ is the channel loss, $e_d$ is the misalignment probability, $d_c$ and $\eta_c$ are the dark count rate and the detection efficiency per detector at the UTP's side, respectively.}
\begin{center}
\includegraphics{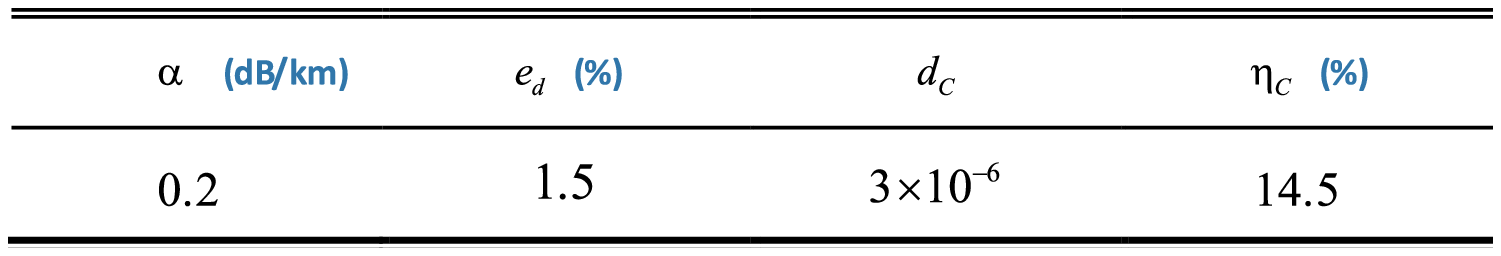}
\end{center}
\end{table}

\begin{figure}[ptb]
\begin{center}
\includegraphics[scale=0.6]{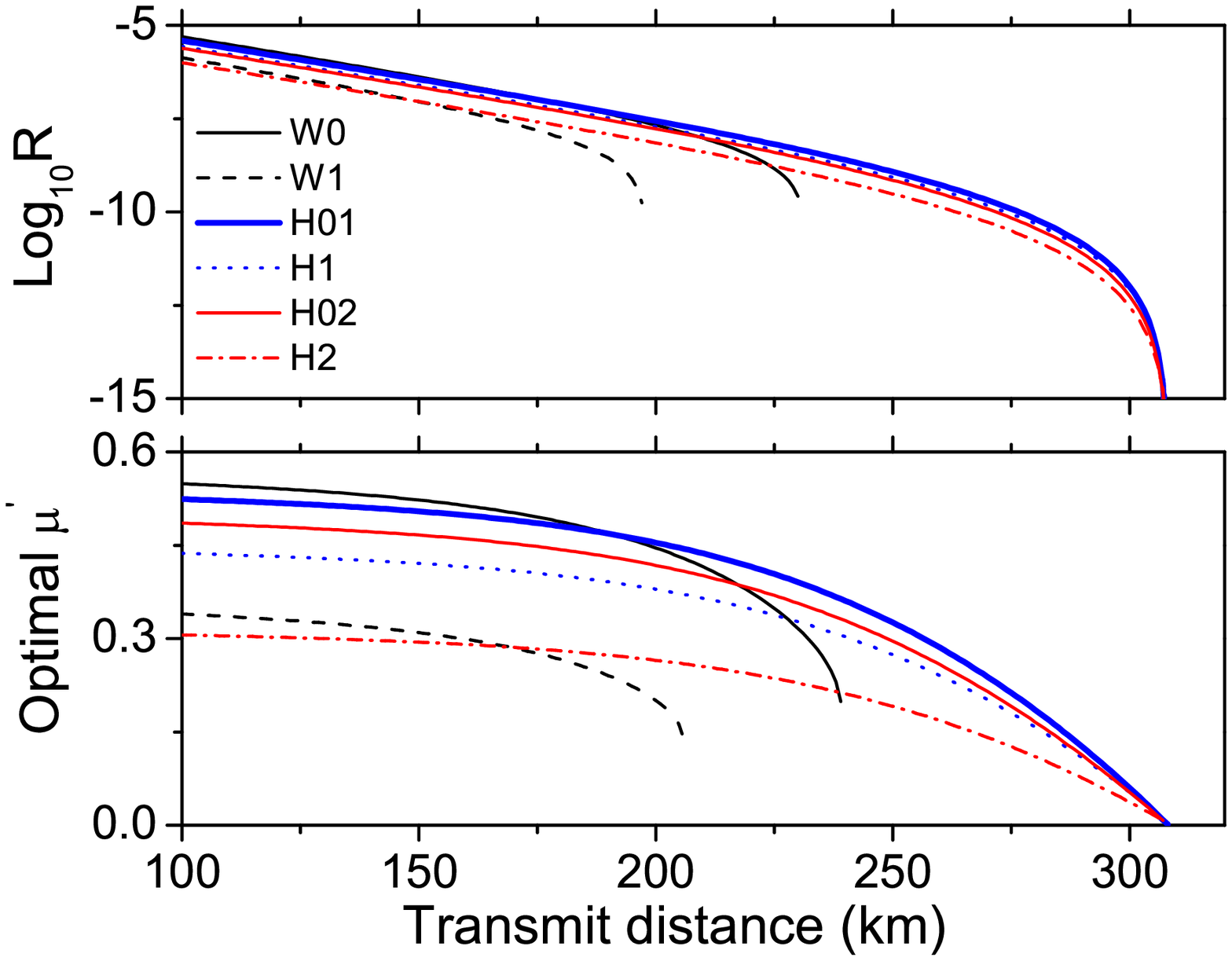}
\end{center}
\caption{(Color online)  (a). Comparison of the final key generation rates vs distance between our proposed scheme and the ones in Ref. \cite{lo2} and Ref. \cite{wang2}.  W0:  infinite intensities with WCS \cite{lo2}; W1: three-intensity method with WCSs; H01 and H02:  infinite intensities with HSPSs;  H1 and H2:  our proposed method with triggered and non-triggered HSPSs. (b). Optimal values of $\mu^{\prime}$ for each curves listed in (a).  The WCSs and the HSPSs used here are all in Poisson distribution. We have chosen the heralding detection efficiency to be 0.9 for curve $H01$ and $H1$, and $0.75$ for curve $H02$ and $H2$.}
\label{Fig2}
\end{figure}

\begin{figure}[ptb]
\begin{center}
\includegraphics[scale=0.6]{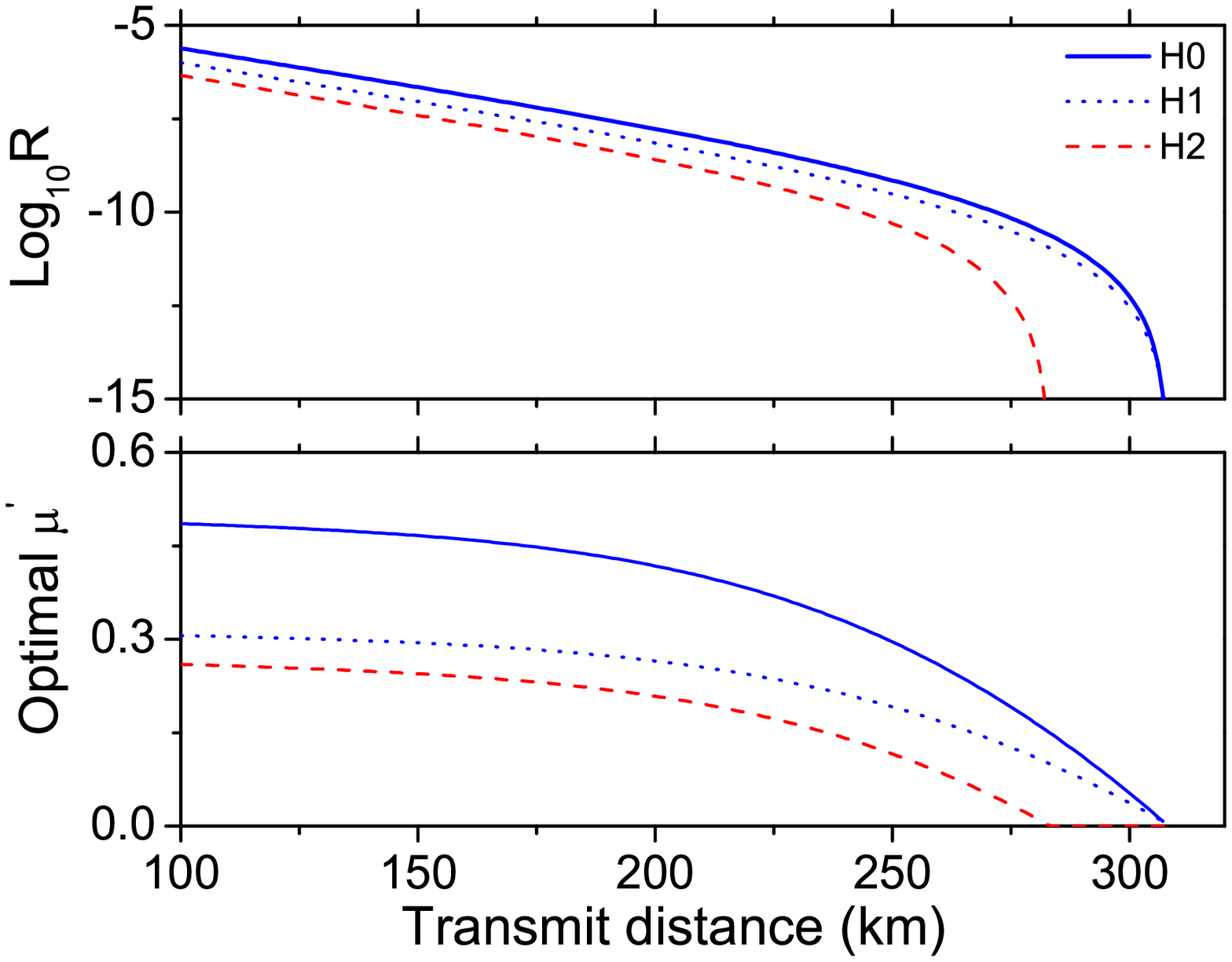}
\end{center}
\caption{(Color online) (a). Comparison of the final key generation rates with HSPSs using different methods.  H0: infinite intensities. H1: a few intensities of this work. H2: key rates of a few intensities using triggered events in sources of intensity $\mu$ and $\mu'$ to calculate $Y_{11}$
\cite{wang2}.   The HSPSs used here are all in poisson distributions. (b). Corresponding optimal values of $\mu^{\prime}$ for all the lines in (a). We have chosen all heralding detection efficiencies to be 0.75.}
\label{Fig3}
\end{figure}

\begin{figure}[ptb]
\begin{center}
\includegraphics[scale=0.6]{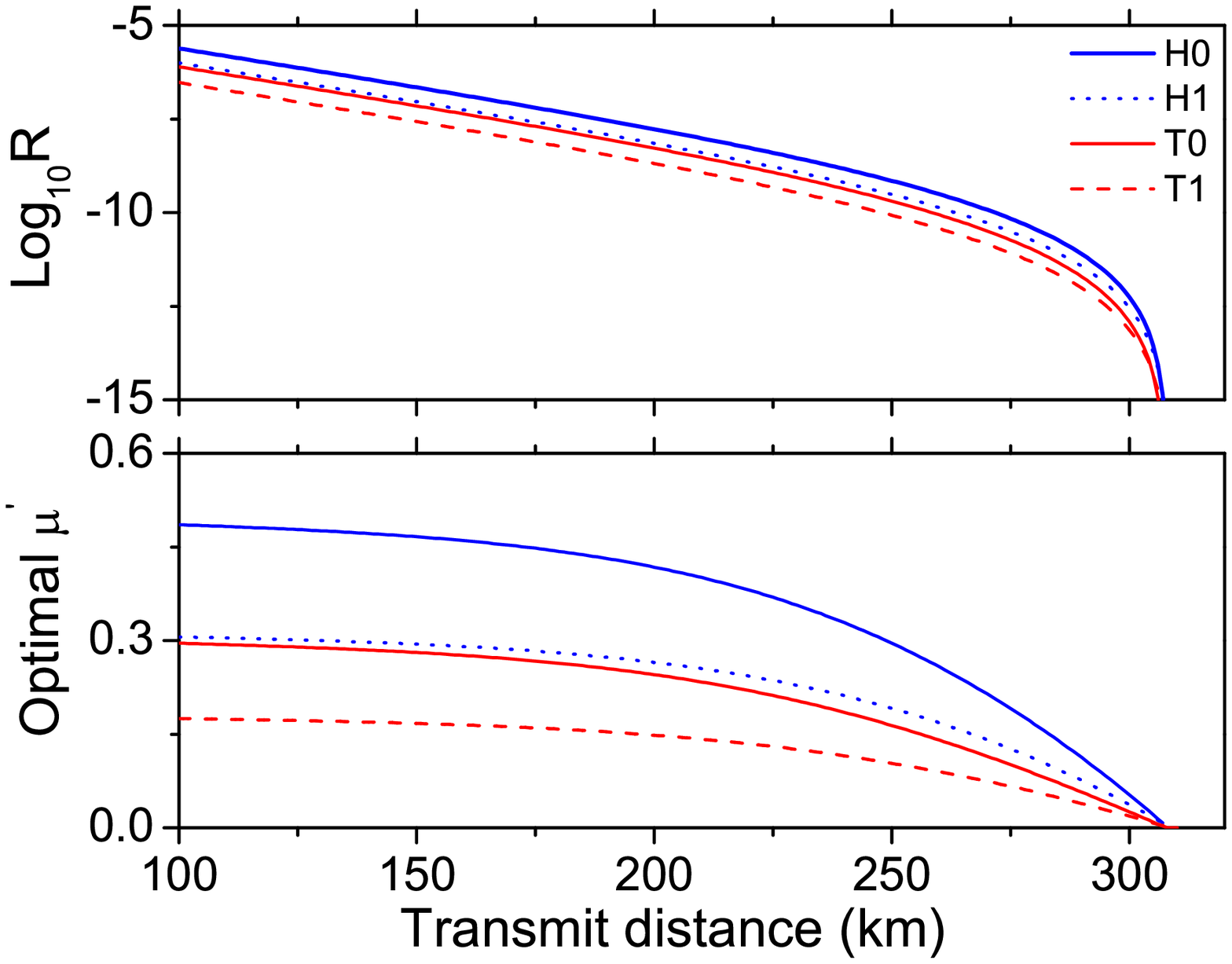}
\end{center}
\caption{(Color online) (a). Comparison of the final key generation rates of MDI-QKD with HSPSs in different photon-number distributions. H0: infinite intensities, Poisson distribution. H1: a few intensities of this work, Poisson distribution; T0: infinite intensities, thermal distribution, T1: a few intensities of this work, thermal distribution. We have chosen all heralding detection efficiencies to be 0.75.}
\label{Fig4}
\end{figure}

Fig. 2(a) displays the comparison of the final key generation rate between different schemes. The curve W0 is the case of using infinite decoy states with WCS \cite{lo2}, W1 represents Wang's three decoy-state method with WCSs \cite{wang2}, H01 (or H02) shows the asymptotic case with HSPSs, and H1 (or H2) represents the result of our new scheme with triggered and non-triggered HSPSs. In the simulations above, we use the optimal values of $\mu^{\prime}$ at each distance for all the curves. Just the difference is: For the asymptotic cases (W0, H01 and H02), the fraction of two-single-photon counts and the QBER of two-single-photon pulses are known exactly; For the normal three decoy-state case (W1), we use the parameters shown in Table I and assume a reasonable value for $\mu$ ($0.1$); While for our new scheme (H1 and H2), we use the same parameters as in Table I except that $\eta_{A}=\eta_{B}=0.9$ (or $0.75$), and borrowing the relationship of $\mu$ and $\mu^{\prime}$ from Eq. (\ref{sc}). Fig. 2(b) shows corresponding optimal values of $\mu^{\prime}$ for each curve in Fig. 2(a). Besides, The WCSs and HSPSs used here are all in poisson distributions.

Fig. 3(a) and (b) are the comparison of our new MDI-QKD scheme with normal three decoy-state method \cite{wang2} using HSPSs. Fig. 3(a) shows the the final key generation rate vs transmission distance, and Fig. 3(b) corresponds to the optimal values of $\mu^{\prime}$. The curve H0 and H1 each corresponds to the asymptotic case with infinite decoy states and our new scheme, respectively. H2 represents the result of using normal three decoy-state method. Here the HSPSs used are all in poisson distributions.

Fig. 4(a) and (b) describe the different behavior of our new MDI-QKD scheme when using HSPSs in different distributions. The curve H0 and H1 each represents the result of using infinite decoy-state method and our new scheme, respectively, and both using HSPSs in poisson distributions. While the lines T0 and T1 correspond to the results of with thermal distributions.

From the comparison above we find that:

(i). Our new scheme of using triggered and non-triggered signals can work excellently close to the asymptotic case with infinite decoy-state method as in Fig 2(a) and (b). This is due to the precise estimation of the tight bounds of $Y_{11}$ and $e_{11}$ by using both triggered and non-triggered signals.

(ii). Our new MDI-QKD scheme with HSPSs can transmit a much longer distance compared with the one with WCSs ($>$ 70km here) as shown in Fig. 2(a), which benefits from the substantial low vacuum components in the heralded signals.

(iii). In our new scheme, the HSPSs in poisson distributions show similar key generation rates as WCSs at short distances, and much higher key rates at long distances as shown in Fig. 2(a). This is attributed to a much higher optimal value of $\mu^{\prime}$ being used as shown in Fig. 2(b).

(iv) According to our calculation here, the protocol using Eq.(\ref{y11}) can have a higher key rate than the one using only triggered events,  as shown in Fig. 3(a), because of a much higher optimal value of $\mu^{\prime}$ being used as shown in Fig. 3(b).

(v) Similar to Ref. \cite{qin2,hwl}, the HSPS source in poisson distribution has better performance than the one in thermal distribution as shown in Fig. 4(a) and (b). This is because the poisson distribution has a higher single-photon probability.

In all our calculations, we did not normalize the triggered or non-triggered states, e.g., Eq.(\ref{xt},\ref{xnt}). Hence the gains and the key rates calculated here are in the unit of the rate averaged over all pumped events of certain intensity in a certain basis. For example, in H/V basis, there are $N_z'$ times that both Alice and Bob used stronger pump lights. Among these events, they obtain $N_{tn}'$ events of triggering at both sides and $n_{tz}'$ times of successful events. Then the the gain in our definition is $n_{tz}'/N_z'$. If we want the key rate averaged over number of triggered states, our results in each figures becomes several times larger, since it should be multiplied by a factor $1/F$ and $F$ is the normalization factor.

\section{Conclusions and discussions}
In summary, we have studied the decoy-state MDI-QKD with heralded single photon source.
We show that this proposed implementation offers a longer transmission distance compared with  existing realization methods. Therefore, it looks promising for practical applications in the future.

\section{ACKNOWLEDGMENTS}
The author-Q. Wang thanks professor Z. Yang and B. Y. Zheng for useful discussion and kind support during the work. We  gratefully acknowledge the financial support from the National High-Tech Program of China through Grants No. 2011AA010800 and No. 2011AA010803, the NSFC through Grants No. 11274178, No. 11174177 and No. 60725416, and 10000-Plan of Shandong province.


\begin{thebibliography}{99}
\bibitem{note} For those non-triggered events, the observed error rate in $X$ basis is
rather large. To avoid this drawback, one can choose to use the triggered events, i.e., Eq.(13) in estimating the errors in $X$ basis and hence the phase-flip rate.
\end{thebibliography}
\end{document}